\documentstyle[12pt]{article}
\begin{document}
\title{\bf Spectral Asymptotics of Eigen-value Problems
with Non-linear Dependence on the Spectral Parameter}

\bigskip\bigskip
\bigskip\bigskip
\bigskip\bigskip
\bigskip\bigskip

\author{       Dmitri V. Fursaev ~\thanks{
        fursaev@thsun1.jinr.ru}  \\ {}\\
     {\small\it Bogoliubov Laboratory of Theoretical Physics}\\
{\small \it  Joint Institute for Nuclear Research}\\
{\small \it  141980 Dubna, Moscow Region, Russia}
\\
}
\maketitle
%%%%%%%%%%%%%%%%%%%%%%%%%%%%%%%%%%%%%%%%%%%%%%%%%%%%%%%
\begin{abstract}
We study asymptotic distribution of eigen-values $\omega$
of a quadratic operator polynomial
of the following form $(\omega^2-L(\omega))\phi_\omega=0$,
where $L(\omega)$ is a second order differential positive elliptic
operator with quadratic dependence on the spectral parameter $\omega$.
We derive asymptotics of the spectral density in this
problem and show how to compute coefficients of its asymptotic expansion
from coefficients of the asymptotic expansion  of the trace of
the heat kernel of $L(\omega)$.
The leading term in the spectral asymptotics is
the same as for a Laplacian in a cavity. The results have
a number of physical applications. We illustrate them by
examples of field equations in external stationary gravitational
and gauge backgrounds.
\end{abstract}
%%%%%%%%%%%%%%%%%%%%%%%%%%%%%%%%%%%%%%%%%%%%%%%%%%

\bigskip
%\vspace{3cm}

\bigskip

\baselineskip=.6cm


\newpage

\section{Introduction}
\setcounter{equation}0

This work is motivated by studying
quantum effects in
stationary gravitational or gauge
classical background fields \cite{Fursaev:2000},
\cite{Fursaev:01}.
If quantum fields $\phi$ are free or one is restricted by
one-loop approximation the fields obey second order
differential equations. In the corresponding
coordinates $(x^0,x^i)$ these equations can be brought to the
following general form:
\begin{equation}\label{1.1}
\left[\partial_0^2+L(i\partial_0)\right]\phi=0.
\end{equation}
$L(i\partial_0)$ is the differential operator which includes both time
derivative $\partial_0=\partial /\partial x^0$ and space derivatives,
\begin{equation}\label{1.2}
L(i\partial_0)=L_2+(i\partial_0)L_1+(i\partial_0)^2L_0,
\end{equation}
$L_k$ are $k$th order differential operators with the space
derivatives only.
This structure of $L(i\partial_0)$ can be inferred by
analyzing dimensionalities.
In what follows we assume that $L_k$ act over a $d$-dimensional
compact space ${\cal M}_d$.
The concrete form of (\ref{1.1}) will be specified
in next section.

In time-independent backgrounds, field excitations
$\phi_\omega(x^0,x^i)=e^{-i\omega x^0}\phi_\omega(x^i)$  with
energy $\omega$ are solutions to
\begin{equation}\label{1.3}
\left[\omega^2-L(\omega)\right]\phi_\omega=0,
\end{equation}
\begin{equation}\label{1.3a}
L(\omega)=L_2+\omega L_1 + \omega^2 L_0.
\end{equation}
Equation (\ref{1.3}) is an eigen-value problem on the spectrum of $\omega$.
If $L_1$ and $L_0$ in (\ref{1.2}) are absent one has a standard problem of
finding eigen-values $\Lambda=\omega^2$ of a second order differential operator
$L_2$.  In general, however, operators $L_1$ and $L_0$ in (\ref{1.3})
are non-trivial and all three operators $L_k$ do not commute.
Thus, (\ref{1.3}) depends
polynomially on the spectral parameter $\omega$.

In this paper
we call (\ref{1.3}) the non-linear spectral problem (NLSP)
(this should not be confused with problems which depend non-linearly
on eigen-functions).
Equations like (\ref{1.3}), (\ref{1.3a})
belong to the spectral theory of polynomial operator
pencils \cite{Markus}, a field of mathematics where
important pioneering results were
established by Keldysh \cite{Keldysh} fifty years ago.
In addition to already mentioned applications,
quadratic and more general operator polynomials
appear in other physical problems, for instance,
oscillations of a viscous fluid,
Schr\"odinger  equation
with energy-dependent potential and etc.

The aim of this work is to analyze asymptotic
distribution of the spectrum
of (\ref{1.3}) at large $\omega$. This can be done by studying
asymptotics  of the spectral density.
If the spectrum is real the information about its properties can be also derived
from the function
\begin{equation}\label{1.4}
K(t)=\frac 12 \sum_\omega e^{-t\omega^2},~~~t>0,
\end{equation}
where $\omega$ are eigen-values  of (\ref{1.3}).
The
behavior of $K(t)$ at small $t$ is connected with
the distribution of large $\omega$.
In next sections we show that at small $t$, if $L(\omega)$ is
a positive elliptic operator,
\begin{equation}\label{1.5}
K(t) \sim {1 \over (4\pi t)^{d/2}}\sum_{n=0}^\infty \left[a_{n} t^{n}+
b_{n} t^{n+1/2}\right].
\end{equation}
This expansion has the same form as asymptotics of the trace of
the heat kernel of a second order differential operator. Moreover,
the coefficient $a_0$ is proportional to the volume of ${\cal M}_d$
while other $a_n$ and $b_n$ are local functionals of the external background
fields
and can be computed by using heat-kernel coefficients of
the operator $L(\omega)$ in (\ref{1.3}).
Thus, the behavior of the spectrum of the NLSP at large $\omega^2$ is
very similar to behavior of the spectrum of a Laplacian in a cavity.

There are several important applications of (\ref{1.5}) in
quantum field theory. For instance, the spectrum of $\omega$ defines
the vacuum energy $E_0$ and, in case of a finite
temperature $\beta^{-1}$, the free-energy of system $F(\beta)$.
For Bose statistics
\begin{equation}\label{appl}
E_0=\frac 12 \sum \omega~~,~~F(\beta)=\beta^{-1}\sum_{\omega}
\ln\left(1-e^{-\beta \omega}\right).
\end{equation}
Asymptotics (\ref{1.5}) determine the form
of divergence of $E_0$ at large
$\omega$. One can also use (\ref{1.5}) to derive the behaviour of
$F(\beta)$ at large temperatures.
For further discussion see \cite{Fursaev:01}.

There are two main approaches to the
operator polynomials discussed in the literature
(see \cite{Markus}): the method of linearization and the method of
factorization. The method of linearization
reduces the spectral problem like (\ref{1.3}) to a standard linear
problem but in an extended Hilbert space. This method is analogous to the
reduction of an $n$th order differential equation to a system of
$n$ first order equations. The factorization method reduces
an operator polynomial to a product of pseudo-differential operators
each of which depends linearly on  the spectral parameter.

The method used in this paper is different. Our idea is to find the spectrum of
(\ref{1.3}) from the spectrum of
$L(\omega)$, by considering at first $\omega$ as an independent parameter.
This way seems to be more simple (at least for discussed class of polynomials)
and it enables us to use
the results of the spectral theory of elliptic operators.
Another new element which is absent in the Keldysh
approach is in using the inner product $<,>$
analogous to the Klein-Gordon product known in relativistic
field theory. This step is motivated
by physical applications and is important for our analysis.
The eigen-functions $\phi_\omega$ with different $\omega$ are orthogonal
with respect to $<,>$ and the set of eigen-functions can be orthonormalized.

Asymptotics (\ref{1.5}) were first reported in our earlier
publication \cite{Fursaev:01} for particular problem
related to fields in stationary but not static backgrounds.
The aim of this work is to make arguments
leading to (\ref{1.5}) more general.

The paper is organized as follows.  In next section we give more detailed
definition of the class of NLSP which will be considered here.
Conditions which restrict this class are dictated
by physical requirements to equation (\ref{1.1}).
We then derive a
relation between the spectral density of eigen-values of (\ref{1.3})
and the spectral density of operators $L(\omega)$. This relation is used in
section 3 in studying spectral asymptotics of the NLSP and for
calculation of coefficients $a_n$ and $b_n$ in (\ref{1.5}). One of the
consequences of these results is that for certain $n$, depending
on the dimension $d$, coefficients $a_n$, $b_n$ are
related to heat kernel coefficients
of the operator $\partial_0^2+L(i\partial_0)$, see
section 4. Section 5 is devoted to examples of NLSP where the
spectrum has a simple form
which allows one a direct
check of the obtained formulas. One example is a charged field in a constant
gauge potential, the other is a field in a rotating
Einstein universe $\mathrm{R}^1\times S^3$. Finally, section 6
is devoted to spectral problems related to the Dirac equation.
Although our results are applicable to these problems additional
consideration is needed because the corresponding
operator $L(\omega)$ is not self-adjoint.

\section{Formulation of the problem}
\setcounter{equation}0

We consider
free fields $\phi$ in a stationary space-time of a dimension $D=d+1$.
Equation (\ref{1.1}) for $\phi$ is reduced to problem (\ref{1.3}),
(\ref{1.3a}) where
$L(\omega)$ will be taken in the following form
\begin{equation}\label{def}
L(\omega)=-(\nabla_k+iA_k+i\omega a_k)(\nabla^k+iA^k+i\omega a^k)
+\omega B+V.
\end{equation}
It will be assumed that $L(\omega)$ acts on sections  to some
vector bundle over ${\cal M}_d$ where ${\cal M}_d$ is a compact
$d$-dimensional
Riemannian manifold with metric $h_{ik}$.
Index $k$ in (\ref{def}) is raised and lowered by using the
metric tensor $h_{ik}$, $\nabla_k$ are
the corresponding
connections on ${\cal M}_d$ and $A_k$, $a_k$, $B$ and $V$ are some
matrix-valued fields.

In what follows we assume that the spectrum of (\ref{1.3})
is real and consider $\omega$ as a real parameter.
We also make two additional assumptions:
i) $L(\omega)$ is a positive
elliptic operator which is self-adjoint on a Hilbert
space $\mathrm{L}^2({\cal M}_d)$
on ${\cal M}_d$ and $L_2=L(0)$ has strictly positive spectrum;
ii) for solutions of (\ref{1.1}) there can be defined
a time-independent product $\langle~,~\rangle$ such that
all positive frequency
eigen-functions $\phi_\omega(x^0,x^i)=e^{-i\omega x^0}
\phi_\omega(x^i)$
$(\omega>0)$ have positive norm
($\langle\phi_\omega,\phi_\omega \rangle>0$) while the norm
of negative-frequency functions $(\omega<0)$ is negative.
According to (i) there are no zero modes ($\omega=0$) in the spectrum.

Positivity of $L(\omega)$ implies that ${\cal M}_d$
is a Euclidean manifold. The second assumption is
dictated by physical considerations.
To quantize the theory the field
$\phi$ is divided onto two parts.
The part which has a positive norm is  connected with
particles while the negative norm part corresponds to
anti-particles.
Our assumption guarantees that energies $\omega$ of
all physical particles are non-negative.

The Hilbert space  $\mathrm{L}^2({\cal M}_d)$ is equipped with
the standard inner product
$(\phi,\psi)=\int \phi^{+}\psi \sqrt{h}d^d x$.
The product $\langle~,~\rangle$ is  defined  as follows
\begin{equation}\label{2.1}
\langle \phi,\psi \rangle=i(\phi,\dot{\psi})-i(\dot{\phi},\psi)-(\phi,L_1\psi)
-i(\phi,L_0\dot{\psi})+i(\dot{\phi},L_0\psi),
\end{equation}
where $\dot{\phi}=\partial_0\phi$ and $L_k$ are Hermitian operators
defined for
(\ref{def}) by (\ref{1.3a}).
One can check that
for any two solutions
$\phi$ and $\psi$ of (\ref{1.1}) which belong to
$\mathrm{L}^2({\cal M}_d)$  (\ref{2.1}) does not depend on $x^0$,
$\partial_0\langle \phi,\psi \rangle=0$.
In a covariant theory on a globally hyperbolic
space-time the latter property is
extended to independence on the choice of a space-like
hyper-surface. For scalar fields (\ref{2.1}) coincides with
the Klein-Gordon product.

Consider the eigen-value problem for operator $L(\omega)$
\begin{equation}\label{2.2}
L(\omega)\phi^{(\omega)}_\Lambda=\Lambda(\omega)\phi^{(\omega)}_\Lambda,
\end{equation}
where $\omega$ is a real parameter.
If the spectrum $\Lambda(\omega)$ is known the spectrum of NLSP (\ref{1.3})
is determined by roots of equation
\begin{equation}\label{2.3}
\chi(\omega,\Lambda)=0,
\end{equation}
\begin{equation}\label{2.4}
\chi(\omega,\Lambda)=\omega^2-\Lambda(\omega).
\end{equation}
This also yields the eigen functions of the NLSP,
$\phi_\omega=C_\omega \phi^{(\omega)}_\Lambda$, where $\omega$ is a root
of (\ref{2.3}) and $C_\omega$ is a normalization coefficient.
(Recall that  (\ref{2.3}) is assumed to have only real roots.)
It is not difficult to see that for eigen-functions of (\ref{1.3})
the following property takes place
\begin{equation}\label{2.5}
\langle \phi_\omega,\psi_\sigma \rangle=\delta_{\omega \sigma}~\chi'(\omega)
(\phi_\omega,\psi_\omega),
\end{equation}
where $\delta_{\omega\sigma}=0$ if $\omega\neq \sigma$ and
$\delta_{\omega\sigma}=1$ if $\omega= \sigma$.
$\chi'(\omega)$ denotes the derivative $\partial_\omega\chi(\omega,\Lambda)$
computed at $\omega$ which are roots of (\ref{2.3}).
It follows from (\ref{2.5}) that any two
eigen-functions of (\ref{1.3}) with different eigen-values
($\omega\neq\sigma$) are automatically orthogonal with respect to $\langle~,~\rangle$;
functions with equal eigen-values are orthogonal
if they are orthogonal with respect to the standard inner
product $(~,~)$.
Thus, the product $\langle~,~\rangle$ can be used
to construct ortho-normalized set of eigen-functions
$\phi_\omega(x^0,x^i)$ for problem (\ref{1.3}).

Because the norm $(\phi_\omega,\phi_\omega)$ is positive
the sign of the norm $\langle\phi_\omega,\phi_\omega\rangle$, according to
(\ref{2.5}), coincides with the sign of $\chi'(\omega)$. Therefore,
condition (ii) is equivalent to the requirement
$\chi'(\omega)=\varepsilon(\omega)|\chi'(\omega)|$,
where $\varepsilon(\omega)$ is the sign function.
Let us also note that verification of (ii) is simplified
when field equations are invariant with respect to the
charge conjugation. In this case the eigen-values $\Lambda(\omega)$ are
symmetric functions of $\omega$.

Consider now function $K(t)$ defined by (\ref{1.4}). By following
\cite{Fursaev:01} we call it {\it pseudo-trace}, keeping in mind that
$K(t)$ may not be a trace of any operator. In particular cases,
when  $L(\omega)=L$,
the pseudo-trace coincides with
the trace of the heat kernel of $L$. This is the reason
why normalization coefficient $1/2$ is introduced in
(\ref{1.4}).
According to definitions (\ref{2.2})--(\ref{2.4})
$K(t)$ can be represented in the integral form
\begin{equation}\label{2.6}
K(t)=\frac 12 \int^{\infty}_{-\infty}d\omega \sum_{\Lambda(\omega)}\delta
(\chi(\omega,\Lambda))~|\partial_\omega\chi(\omega,\Lambda)|~e^{-t\omega^2}.
\end{equation}
By using the integral representation for the Dirac $\delta$-function and
the fact that the signs of $\chi'(\omega)$ and
$\omega$ coincide (\ref{2.6}) can be transformed further
$$
K(t)={1 \over 4\pi}
\int^{\infty}_{-\infty}d\omega~
\varepsilon(\omega)\sum_{\Lambda(\omega)}\int_{-\infty}^{\infty}
dx~ e^{ix\chi(\omega,\Lambda)}
~\partial_\omega\chi(\omega,\Lambda)~e^{-t\omega^2}
$$
\begin{equation}\label{2.7}
={1 \over 4\pi}
\int^{\infty}_{-\infty}d\omega~
\varepsilon(\omega)\int_{C}
dz ~e^{-\omega^2(t-iz)}
\left(2\omega+{1 \over iz}\partial_\omega\right)K_\omega(iz),
\end{equation}
where
$K_\omega(t)$ is the trace of the heat kernel of the operator
$L(\omega)$
\begin{equation}\label{2.8}
K_\omega(t)=\mbox{Tr}~e^{-tL(\omega)}=\sum_{\Lambda(\omega)}e^{-t\Lambda(\omega)}.
\end{equation}
The integration over $x$ in (\ref{2.7}) is shifted  to the complex plane.
Contour $C$
goes from $-i\epsilon-\infty$ to $-i\epsilon+\infty$ where $\epsilon$ is a small positive
parameter.
Equation (\ref{2.7}) enables one to relate the pseudo-trace to the
trace of an elliptic operator.
To proceed we represent $K_\omega(t)$ in the integral form
\begin{equation}\label{2.9}
K_\omega(t)=\int_{\mu}^{\infty} e^{-t\lambda} \varphi(\lambda,\omega)
d\lambda,
\end{equation}
where $\varphi(\lambda,\omega)$ is the {\it spectral density},
which can be written as the sum
of delta-functions $\delta(\lambda-\Lambda(\omega))$. Parameter
$\mu$ is chosen to
be smaller than the lowest eigen-value\footnote{Because
$L(\omega)$ is a positive elliptic operator in a finite volume
its spectrum is bounded from below.}
$\Lambda_0(\omega)$. In principle, the choice of $\mu$ depends on
$\omega$. If the spectrum $\Lambda(\omega)$ is strictly
positive it is convenient to put $\mu =0$.
Let us also introduce the {\it counting function}
of $L(\omega)$
\begin{equation}\label{2.10}
N(\lambda,\omega)=\int_{\mu}^\lambda d\sigma \varphi(\sigma,\omega),
\end{equation}
which is equal to
the total number of eigen-values $\Lambda(\omega)$ which do not exceed $\lambda$.
We use now (\ref{2.9}) in (\ref{2.7}) and take into account (\ref{2.10})
$$
K(t)
$$
\begin{equation}\label{2.11}
={1 \over 4\pi} \int^{\infty}_{-\infty}d\omega~
\varepsilon(\omega)\int_{C}
dz ~e^{-\omega^2(t-iz)}\int_{\mu}^\infty d\lambda e^{-iz\lambda}
\left(2\omega \varphi(\lambda,\omega)+
{1 \over iz}\partial_\omega\partial_\lambda N(\lambda,\omega)\right).
\end{equation}
The last term in the brackets can be integrated by parts over $\lambda$.
Then the integral over $z$ results in the delta-function
$\delta(\lambda-\omega^2)$ and we get the final expression
\begin{equation}\label{2.12}
K(t)=\int_{0}^{\infty} d\lambda~e^{-\lambda t}\varphi(\lambda),
\end{equation}
\begin{equation}\label{2.13}
\varphi(\lambda)=\frac 12 \left(\tilde{\varphi}(\lambda,\sqrt{\lambda})+
\tilde{\varphi}(\lambda,-\sqrt{\lambda})\right),
\end{equation}
\begin{equation}\label{2.14}
\tilde{\varphi}(\lambda,\omega)=\varphi(\lambda,\omega)+
{1 \over 2\omega}\partial_\omega N(\lambda,\omega).
\end{equation}
Note that (\ref{2.12}) does not depend on $\mu$.
Representation (\ref{2.12}) \footnote{Formula (\ref{2.12}) can be also
written as $K(t)=\int_{0}^{\infty} ~d\omega~ e^{-t\omega^2}\Phi(\omega)$
where $\Phi(\omega)=2\omega~\varphi(\omega^2)$. This definition is used
in \cite{Fursaev:01}.}
is convenient because it has the same form
as integral representation (\ref{2.9}) for an ordinary elliptic operator.
Together with (\ref{2.13}), (\ref{2.14}) it will be our basic
equation.

\section{Short $t$ expansions and spectral asymptotics}
\setcounter{equation}0

We begin with short $t$ expansion of the trace $K_\omega(t)$, which is
standard,
\begin{equation}\label{3.3}
K_\omega(t) \sim {1 \over (4\pi t)^{d/2}}\sum_{n=0}^\infty
\left[a_{n}(\omega)t^{n} +b_{n}(\omega)t^{n+1/2}\right],
\end{equation}
where $a_{n}(\omega)$ and $b_n(\omega)$ are
Hadamard--Minackshisundaram--DeWitt--Seeley coefficients.
It follows from (\ref{1.2}), (\ref{def}) that
\begin{equation}\label{3.3b}
a_{n}(\omega)=\sum_{m=0}^{n}a_{m,n}\omega^m,~~~
b_{n}(\omega)=\sum_{m=0}^{n}b_{m,n}\omega^m.
\end{equation}
The highest power of $\omega$ in $a_n(\omega)$ and $b_n(\omega)$
can be found by using (\ref{def}) and analyzing dimensionalities.
It is easy to see that this power
is determined by
the term $\omega B$ in $L(\omega)$, see (\ref{def}).

We are interested in (\ref{3.3}) because
short $t$ limit is related to distribution
of large eigen-values $\Lambda$ in (\ref{2.8}). For instance,
the first term
in series (\ref{3.3}) corresponds to the leading asymptotics of
the  spectral function $N(\lambda,\omega)$ \cite{Titchmarsh}
\begin{equation}\label{3.19a}
N(\lambda,\omega)\sim {\lambda^{d/2}~ a_0 \over (4\pi)^{d/2} \Gamma(d/2+1)}
\sim r{\lambda^{d/2}~ V_d \over (4\pi)^{d/2} \Gamma(d/2+1)},
\end{equation}
where $V_d$ is the volume of ${\cal M}_d$ and $r$ is the
dimensionality of the representation of the field $\phi$.
Equation (\ref{3.19a}) is known as the Weyl formula.
It is impossible, however, to define all other
sub-leading terms in $N(\lambda,\omega)$ corresponding to
(\ref{3.3}). The reason is that
starting with certain $n$
these terms become  smaller
then fluctuations of $N(\lambda,\omega)$ when $\lambda$ goes from
one eigen-value to the next one \cite{Bloch}. The way
out of this difficulty is to work with smoothed functions
$N(\lambda,\omega)$ and $\varphi(\lambda,\omega)$,
see, e.g., \cite{Bloch},\cite{Fulling:82}.

The smoothing can be done by different ways and
one of them is to use the Riesz means \cite{Fulling:99}.
Let us define a "smoothed" spectral function
$\varphi_\alpha(\lambda,\omega)$ by formula
\begin{equation}\label{3.1}
K_\omega(t)t^{-\alpha}=\int_{\mu}^{\infty} e^{-t\lambda} \varphi_\alpha (\lambda,\omega)
d\lambda,
\end{equation}
where $\alpha$ is a complex parameter, $\mathrm{Re}~\alpha >0$.
In particular,
$\varphi_{1}(\lambda,\omega)=N(\lambda,\omega)$ and
in the limit $\alpha\rightarrow 0$
$\varphi_{0}(\lambda,\omega)=\varphi(\lambda,\omega)$.
By using the inverse Laplace
transform in (\ref{3.1}) one gets
\begin{equation}\label{3.2}
\varphi_\alpha(\lambda,\omega)={1 \over
\Gamma(\alpha)}\int_{\mu}^{\lambda}
(\lambda-\sigma)^{\alpha-1}\varphi(\sigma,\omega) d\sigma.
\end{equation}
In fact,
$\varphi_\alpha(\lambda,\omega)$ coincides with the fractional
derivative
$\partial_\lambda^{-\alpha}\varphi(\lambda,\omega)$ of $\varphi(\lambda,\omega)$
of the order $-\alpha$, see \cite{frac}.
Suppose that $\alpha\neq k/2$ where $k$ is an integer.
Then at large $\lambda$ the spectral
function $\varphi_\alpha(\lambda,\omega)$ is represented by the asymptotic
series
\begin{equation}\label{3.4}
\varphi_\alpha(\lambda,\omega) \sim {1
\over (4\pi)^{d/2}}\sum_{n=0}^\infty\left[
a_{n}(\omega)~ { \lambda^{d/2-n+\alpha-1}
\over \Gamma\left(\frac d2 -n +\alpha\right)}
+b_{n}(\omega)~ { \lambda^{(d-1)/2-n+\alpha-1}
\over \Gamma\left({d-1 \over 2} -n +\alpha\right)}
\right].
\end{equation}
If $\mu=0$ one can substitute (\ref{3.4}) in (\ref{3.1}) and check that this series
corresponds to the short $t$ expansion (\ref{3.3}). This result is valid
also if $\mu<0$ at $\lambda>|\mu|$, see Appendix.

The difficulty which appears when one wants to use (\ref{3.4})
to get asymptotics
of $\varphi(\lambda,\omega)$ or $N(\lambda,\omega)$ is that
some terms in (\ref{3.4}) disappear in the
limit $\alpha=0$ or $\alpha=1$. Thus,
these expansions cannot reproduce the entire series (\ref{3.3}).
It was recently pointed out by Dowker \cite{Dowker:01} that for $\mu=0$
the problem can be formally avoided
if in this limit the sub-leading
terms are treated as generalized functions\footnote{Another possible way to
avoid this problem is to use a dimensional regularization
\cite{Fursaev:2000}.}. The recipe of \cite{Dowker:01}
uses the fact that \cite{GS}
\begin{equation}\label{3.5}
\lim_{\beta\rightarrow -n}{x_{+}^{\beta-1} \over
\Gamma(\beta)}=\partial_x^{n}\delta(x),
\end{equation}
where $n=0,1,2,...$ and $x_{+}^{\beta-1}=x^{\beta-1}$ for $x\geq 0$ and
$x_+^{\beta-1}=0$ for $x<0$.
We will use this approach in what follows.

To get
asymptotics  of the spectral density $\varphi(\lambda)$ of
the pseudo-trace $K(t)$ we first define the functions
\begin{equation}\label{3.7}
\tilde{\varphi}_\alpha(\lambda,\omega)=\varphi_\alpha(\omega,\lambda)
+{1 \over 2\omega}\partial_\omega\varphi_{\alpha+1}(\lambda,\omega)
\end{equation}
for $\alpha$ complex. Then
$\tilde{\varphi}(\lambda,\omega)$ introduced in
(\ref{2.14}) can be  obtained from
$\tilde{\varphi}_\alpha(\lambda,\omega)$
in the limit $\alpha\rightarrow 0$.
It follows from (\ref{3.4}) that at large $\lambda$
$$
\frac 12\left(\tilde{\varphi}_\alpha(\lambda,\omega)+
\tilde{\varphi}_\alpha(\lambda,-\omega)\right)
$$
\begin{equation}\label{3.6}
\sim {1
\over (4\pi)^{d/2}}\sum_{n=0}^\infty
\left[
\tilde{a}_{n}(\omega)~ { \lambda^{d/2-n+\alpha-1}
\over \Gamma\left(\frac d2 -n +\alpha\right)}
+\tilde{b}_{n}(\omega)~ { \lambda^{(d-1)/2-n+\alpha-1}
\over \Gamma\left({d-1 \over 2} -n +\alpha\right)}
\right],
\end{equation}
\begin{equation}\label{3.8}
\tilde{a}_{n}(\omega)=\frac 12\left[a_{n}(\omega)+a_{n}(-\omega)+
{1 \over 2\omega}\partial_\omega(a_{n+1}(\omega)+a_{n+1}(-\omega))\right],
\end{equation}
\begin{equation}\label{3.9}
\tilde{b}_{n}(\omega)=\frac 12 \left[b_{n}(\omega)+b_{n}(-\omega)+
{1 \over 2\omega}\partial_\omega(b_{n+1}(\omega)+b_{n+1}(-\omega))\right],
\end{equation}
where we have taken into account that $\partial_\omega a_0(\omega)=\partial_\omega
b_0(\omega)=0$. According to (\ref{3.3b}) one can write
\begin{equation}\label{3.10}
\tilde{a}_{n}(\omega)=\sum_{m=0}^{\infty}\tilde{a}_{2m,n}\omega^{2m},~~~
\tilde{b}_{n}(\omega)=\sum_{m=0}^{\infty}\tilde{b}_{2m,n}\omega^{2m},
\end{equation}
\begin{equation}\label{3.11}
\tilde{a}_{2m,n}=a_{2m,n}+(m+1)a_{2(m+1),n+1},~~
\tilde{b}_{2m,n}=b_{2m,n}+(m+1)b_{2(m+1),n+1},
\end{equation}
where $a_{2m,n}$, $b_{2m,n}$ are assumed to be equal to zero
for $2m>n$. Then
$$
\frac 12 \left(\tilde{\varphi}_\alpha(\lambda,\sqrt{\lambda})+
\tilde{\varphi}_\alpha(\lambda,-\sqrt{\lambda})\right)
$$
\begin{equation}\label{3.12}
\sim {1\over (4\pi)^{d/2}}\sum_{n=0}^\infty\left[
a_{n}^{(\alpha)}~ { \lambda^{d/2-n+\alpha-1}
\over \Gamma\left(\frac d2 -n +\alpha\right)}
+b_{n}^{(\alpha)}~ { \lambda^{(d-1)/2-n+\alpha-1}
\over \Gamma\left({d-1 \over 2} -n +\alpha\right)}
\right].
\end{equation}
Coefficients $a_n^{(\alpha)}$, $b_n^{(\alpha)}$ can be found
with the help of (\ref{3.10}), (\ref{3.11}). After some algebra one
can represent them in the form
\begin{equation}\label{3.13}
a_n^{(\alpha)}=\sum_{m=n}^{2n}(-1)^{n-m}{\Gamma\left(-\frac d2+m-\alpha\right)
\over \Gamma\left(-\frac d2+n-\alpha\right) } a_{2(m-n),m},
\end{equation}
\begin{equation}\label{3.14}
b_n^{(\alpha)}=\sum_{m=n}^{2n}(-1)^{n-m}{\Gamma\left(-{d-1 \over 2}+m-\alpha\right)
\over \Gamma\left(-{d-1 \over 2}+n-\alpha\right) } b_{2(m-n),m}.
\end{equation}
Finally, asymptotics of the spectral density (\ref{2.13}) is obtained from
(\ref{3.12}) in the limit $\alpha\rightarrow 0$ by treating its coefficients
as generalized functions, see (\ref{3.7}),
$$
\varphi(\lambda)=\frac 12 \lim_{\alpha\rightarrow 0}
\left(\tilde{\varphi}_\alpha(\lambda,\sqrt{\lambda})+
\tilde{\varphi}_\alpha(\lambda,-\sqrt{\lambda})\right)
$$
\begin{equation}\label{3.15}
\sim {1\over (4\pi)^{d/2}}\sum_{n=0}^\infty\left[
a_{n}~ \partial_\lambda^{n-d/2}
+b_{n}~\partial_\lambda^{n-(d-1)/2}\right]\delta(\lambda),
\end{equation}
where $a_n=a_n^{(0)}$, $b_n=b_n^{(0)}$ and the symbol
$\partial_\lambda^{\gamma}$ for $\gamma\neq n$ denotes
the fractional derivative.
Note that $L(\omega)$ may have negative eigen-values at some $\omega$.
In this case there is the restriction $\lambda>|\mu|>0$ in (\ref{3.4})
where $\mu=\mu(\omega)$ is determined by the lowest eigen-value $\Lambda_0(\omega)$
of $L(\omega)$,
$\mu(\omega)<\Lambda_0(\omega)$.
In (\ref{3.12}) this restriction is absent because
the lowest eigen-value and $\mu$ depend on parameter $\lambda$.
When $\lambda$ decreases and goes to zero so does $|\mu|$  (recall that spectrum
of $L_2=L(0)$ is strictly positive).
Therefore, using formula (\ref{3.5}) in (\ref{3.12}) is legitimate.

Equation (\ref{3.15}) can be used to get the normal asymptotic series
for a smoothed spectral density at large $\lambda$. For the Riesz means
with $\mathrm{Re}~\alpha > 0$, $\alpha\neq k/2$,
\begin{equation}\label{3.2a}
\varphi_\alpha(\lambda)={1 \over
\Gamma(\alpha)}\int_{0}^{\lambda}
(\lambda-\sigma)^{\alpha-1}\varphi(\sigma) d\sigma,
\end{equation}
\begin{equation}\label{3.12a}
\varphi_\alpha(\lambda)\sim {1\over (4\pi)^{d/2}}\sum_{n=0}^\infty\left[
a_{n}~ { \lambda^{d/2-n+\alpha-1}
\over \Gamma\left(\frac d2 -n +\alpha\right)}
+b_{n}~ { \lambda^{(d-1)/2-n+\alpha-1}
\over \Gamma\left({d-1 \over 2} -n +\alpha\right)}
\right],
\end{equation}
\begin{equation}\label{3.17}
a_n=\sum_{m=n}^{2n}(-1)^{n-m}{\Gamma\left(-\frac d2+m\right)
\over \Gamma\left(-\frac d2+n\right) } a_{2(m-n),m},
\end{equation}
\begin{equation}\label{3.18}
b_n=\sum_{m=n}^{2n}(-1)^{n-m}{\Gamma\left(-{d-1 \over 2}+m\right)
\over \Gamma\left(-{d-1 \over 2}+n\right) } b_{2(m-n),m}.
\end{equation}
The equivalent and regularization independent
way to represent this result is the
short $t$ expansion of the pseudo-trace
$K(t)$ which
is the direct consequence of (\ref{2.12}) and (\ref{3.15})
\begin{equation}\label{3.16}
K(t) \sim {1 \over (4\pi t)^{d/2}}\sum_{n=0}^\infty
\left[a_{n}t^{n} +b_{n}t^{n+1/2}\right].
\end{equation}
Brief comments about (\ref{3.12a}) and (\ref{3.16})
are in order.
1) It is important fact that
$a_n$ and $b_n$ are finite combinations of some pieces,
$a_{2m,n}$, $b_{2m,n}$, of the heat kernel coefficients
of the associated operator polynomial $L(\omega)$, see (\ref{def}), (\ref{3.3}),
(\ref{3.3b}). Thus, studying the spectral asymptotics
of the considered class of non-linear spectral problems is reduced
to standard computations.
2) One can use
(\ref{3.12a}) to get the leading asymptotics of the counting
function $N(\lambda)=\varphi_1(\lambda)$ for the given class of NLSP
\begin{equation}\label{3.19}
N(\lambda)\sim r{\lambda^{d/2}~ V_d \over (4\pi)^{d/2} \Gamma(d/2+1)},
\end{equation}
This equation follows from (\ref{3.17})
and the fact that $a_{0,0}=a_0=rV_d$.
It means that behavior of
$N(\lambda)$ is described by the Weyl formula.
The difference between
NLSP spectrum and spectrum of a Laplacian appears only in the
sub-leading corrections to (\ref{3.19}).
3) The form
of asymptotics (\ref{3.12a}), (\ref{3.16})
strongly depends on the form
of operator polynomial.  The heat kernel coefficients for
$L(\omega)$ defined by (\ref{def})
are polynomials (\ref{3.3b}) of the certain order which guarantees
that $a_n$, $b_n$
are finite series given by (\ref{3.17}), (\ref{3.18}).
These formulas may not hold for other operator polynomials.
The leading term in spectral asymptotics
for some class of operator polynomials
is discussed also in \cite{Markus}.

\section{Dimensional reduction}
\setcounter{equation}0

Spectral problem (\ref{1.3})
is related to wave equation (\ref{1.1}) in a stationary $d+1$ dimensional
Lorentzian space-time
${\cal M}_{d+1}$. One can  introduce a differential operator $P$ acting over
${\cal M}_{d+1}$
\begin{equation}\label{4.1}
P=\partial_0^2+L(i\partial_0),
\end{equation}
where $L(i\partial_0)$ is defined in (\ref{1.2}).
The heat kernel of $P$ and its asymptotics \cite{DW}
\begin{equation}\label{4.2}
K_{d+1}(s)=\mbox{Tr} ~ e^{-s P}\sim {1 \over i(4\pi s)^{(d+1)/2}}\sum_{n=0}^\infty
\left[A_n s^n +B_n s^{n+1/2}\right]
\end{equation}
at small $|s|$ play an important
role in quantum field theory on ${\cal M}_{d+1}$. (The parameter $s$ in (\ref{4.2})
is imaginary because $P$ is a hyperbolic operator  \cite{DW}
and, strictly speaking, $K_{d+1}(s)$ should be solution to a Schr\"odinger problem.)
There is a relation between asymptotics (\ref{4.2})
and asymptotics of the related NLSP.
To see this let us first
regularize integrals over time $x^0$ by requiring that
$-T/2<x^0<T/2$, where $T$ is a large parameter.
Then we can write
\begin{equation}\label{4.3}
K_{d+1}(s)={T \over 2\pi}\int_{-\infty}^\infty d\omega \sum_{\Lambda(\omega)}
e^{-s(-\omega^2+\Lambda(\omega))}=
{T \over 2\pi}\int_0^\infty d\omega~
e^{s\omega^2}~\left[K_\omega(s)+K_{-\omega}(s)\right].
\end{equation}
At small $|s|$ we can use (\ref{3.3}), (\ref{3.3b}) to get
\begin{equation}\label{4.4}
K_{d+1}(s) \sim {T \over \sqrt{\pi}i(4\pi s)^{(d+1)/2}}\sum_{n=0}^\infty
\sum_{m=0}^{[n/2]}(-1)^m \Gamma\left(m+\frac 12\right)
\left(a_{2m,n}s^{n-m}+b_{2m,n}s^{n-m+1/2}\right).
\end{equation}
By comparing (\ref{4.4}) with (\ref{4.2})
one finds
\begin{equation}\label{4.5}
A_n={T \over \sqrt{\pi}}\sum_{m=n}^{2n}(-1)^{n-m} \Gamma\left(m-n+\frac 12\right)
a_{2(m-n),m},
\end{equation}
\begin{equation}\label{4.6}
B_n={T \over \sqrt{\pi}}\sum_{m=n}^{2n}(-1)^{n-m} \Gamma\left(m-n+\frac 12\right)
b_{2(m-n),m}.
\end{equation}
These formulas yield heat kernel coefficients in $d+1$
dimensions
in terms of some pieces of coefficients in $d$
dimensions.
Relation
to coefficients of the pseudo-trace expansion
follows if (\ref{4.5}), (\ref{4.6}) are compared with (\ref{3.17}),
(\ref{3.18}), respectively.
For theories in
space-times with even dimensions, $D=d+1=2k$,
\begin{equation}\label{4.7}
A_{D/2}=Ta_{D/2}=\int dt ~a_{D/2},
\end{equation}
for odd dimensions, $D=d+1=2k+1$,
\begin{equation}\label{4.8}
B_{(D-1)/2}=Tb_{(D-1)/2}=\int dt ~b_{(D-1)/2}.
\end{equation}
Relation  (\ref{4.7}) is interesting
because coefficient $A_{D/2}$ in even dimensions determines
anomalous scaling of the
theory. It can be shown that in conformally invariant models it coincides
with the integral of the conformal anomaly (trace of renormalized stress
energy tensor).
An analog of formulas (\ref{4.5})--(\ref{4.8}) can be also obtained in the Euclidean
theory \cite{FZ:01} where
$P$ is an elliptic operator and its heat kernel is
well-defined.

\section{Examples}
\subsection{Constant gauge potential}
\setcounter{equation}0

We now consider concrete examples  which illustrate
results (\ref{3.17})--(\ref{3.16}). These examples
can be obtained by transforming
standard eigen-value problems to non-linear ones and
the spectrum of (\ref{1.3}) can be found explicitly.
Consider the problem
\begin{equation}\label{5.1}
\left[\omega^2-L_2\right]\phi_\omega=0,
\end{equation}
where $L_2$ is a second order positive elliptic
operator on a compact space. We
suppose that the lowest eigen-value $\Lambda_0$ of $L_2$ is positive,
$\Lambda_0>0$.
Now, if $\omega$ is replaced to $\omega-\varrho$, where $\varrho$ is
a real parameter,
the eigen-value
problem becomes non-linear
\begin{equation}\label{5.2}
\left[\omega^2-L(\omega)\right]\phi'_\omega=0,
\end{equation}
\begin{equation}\label{5.3}
L(\omega)=L_2-\varrho^2+2\varrho \omega,
\end{equation}
where $\phi'_{\omega}=\phi_{\omega+\varrho}$.
From the point of view of the theory in space and time the shift of
frequencies corresponds to a gauge-like
transformation $\phi_\omega'(x^0,x)=e^{-i\varrho x^0}\phi_\omega(x^0,x)$.
Equations for $\phi_\omega'(x^0,x)$ look as equations in external
constant gauge potential $A_\nu dx^\nu=\varrho dx^0$.
Another application of these results is a finite-temperature theory
where $\varrho$ plays the role of a chemical potential.
In what follows, to satisfy condition (ii)
of section 2, we assume that $\varrho^2<\Lambda_0$.

The pseudo-trace corresponding to (\ref{5.2}) is
\begin{equation}\label{5.4}
K(t)=\frac 12\sum_{\omega}e^{-t\omega^2}=
\frac 12\sum_{\Lambda}\left(e^{-t(\sqrt{\Lambda}-\varrho)^2}+
e^{-t(\sqrt{\Lambda}+\varrho)^2}\right),
\end{equation}
where $\Lambda$ are eigen-values of $L_2$.
One can consider (\ref{5.4}) as a result of a non-linear transformation
of the spectrum, $\Lambda\rightarrow (\sqrt{\Lambda}\pm\varrho)^2$.
The spectral density $\varphi(\lambda)$ is defined by (\ref{2.12})
and can be written as
\begin{equation}\label{5.6}
\varphi(\lambda)={1 \over 2\sqrt{\lambda}}
\left[|\sqrt{\lambda}+\varrho|~
\bar{\varphi}\left((\sqrt{\lambda}+\varrho)^2\right)+
|\sqrt{\lambda}-\varrho|~
\bar{\varphi}\left((\sqrt{\lambda}-\varrho)^2\right)\right],
\end{equation}
where $\bar{\varphi}(\lambda)$ is the spectral density of
$L_2$.
The asymptotic behavior of the smoothed density
$\bar{\varphi}_\alpha(\lambda)=
\partial^{-\alpha}_\lambda\bar{\varphi}(\lambda)$,
$\mathrm{Re}~\alpha>0$, $\alpha \neq k/2$, is
\begin{equation}\label{5.7}
\bar{\varphi}_\alpha(\lambda) \sim {1
\over (4\pi)^{d/2}}\sum_{n=0}^\infty\left[
\bar{a}_{n}~ { \lambda^{d/2-n+\alpha-1}
\over \Gamma\left(\frac d2 -n +\alpha\right)}+
\bar{b}_{n}~ { \lambda^{(d-1)/2-n+\alpha-1}
\over \Gamma\left({d-1\over 2} -n +\alpha\right)}
\right],
\end{equation}
where $\bar{a}_{n}$ are heat kernel coefficients of  $L_2$
\begin{equation}\label{5.7a}
\mbox{Tr}~e^{-tL_2}\sim {1 \over (4\pi t)^{d/2}}\sum_{n=0}^{\infty} \left[\bar{a}_n
t^n+\bar{b}_nt^{n+1/2}\right].
\end{equation}
By taking into account that
\begin{equation}\label{form}
(x-y)^z=\sum_{p=0}^\infty {\Gamma(-z+p) \over p! \Gamma(-z)} x^{z-p} y^p,
\end{equation}
$|y|<x$, one finds from (\ref{5.7}) at large $\lambda$
$$
{1 \over 2\sqrt{\lambda}}
\left[|\sqrt{\lambda}+\varrho|~
\bar{\varphi}_\alpha\left((\sqrt{\lambda}+\varrho)^2\right)+
|\sqrt{\lambda}-\varrho|~
\bar{\varphi}_\alpha
\left((\sqrt{\lambda}-\varrho)^2\right)\right]
$$
\begin{equation}\label{5.8}
\sim {1
\over (4\pi)^{d/2}}\sum_{n=0}^\infty
\left[a_{n-\alpha}~ { \lambda^{d/2-n+\alpha-1}
\over \Gamma\left(\frac d2 -n +\alpha\right)}
+b_{n-\alpha}~ { \lambda^{(d-1)/2-n+\alpha-1}
\over \Gamma\left({d-1 \over 2} -n +\alpha\right)}\right],
\end{equation}
\begin{equation}\label{5.9}
a_{n-\alpha}=\sum_{k=0}^n~\varrho^{2k}~ c_{k,n-\alpha}(d)~\bar{a}_{n-k},~~~
b_{n-\alpha}=\sum_{k=0}^n~\varrho^{2k}~ c_{k,n-\alpha}(d-1)~\bar{b}_{n-k},
\end{equation}
\begin{equation}\label{5.10}
c_{k,n-\alpha}(d)={2^{2k} \over (2k)!}~{\Gamma\left({d+1 \over 2} +\alpha-n+k\right)
\over \Gamma\left({d+1 \over 2} +\alpha-n\right)}.
\end{equation}
In the limit $\alpha\rightarrow 0$ one obtains from
(\ref{5.8})--(\ref{5.10}) asymptotics (\ref{3.15})
for $\varphi(\lambda)$, where $a_n=\lim_{\alpha\rightarrow 0}
a_{n-\alpha}$ and  $b_n=\lim_{\alpha\rightarrow 0}
b_{n-\alpha}$.
By using this
one comes to short $t$ expansion (\ref{1.5})
for pseudo-trace (\ref{5.4}).

It is possible to check that obtained formulas (\ref{3.17}), (\ref{3.18})
correctly reproduce $a_n$ and $b_n$.
To this aim one needs to know
coefficients $a_{2m,n}$ in (\ref{3.17}) and $b_{2m,n}$ in  (\ref{3.18}).
They are defined by heat kernel coefficients of the operator $L(\omega)$,
Eq. (\ref{5.3}),
and can be easily found. By using (\ref{5.3})
one gets after some algebra
\begin{equation}\label{5.14}
a_{2m,n}={2^{2m} \over (2m)!} \sum_{p=2m}^n {\varrho^{2p-2m} \over (p-2m)!}
\bar{a}_{n-p},~~~b_{2m,n}={2^{2m} \over (2m)!} \sum_{p=2m}^n {\varrho^{2p-2m} \over (p-2m)!}
\bar{b}_{n-p}
\end{equation}
Combination of (\ref{3.17}), (\ref{3.18}) with (\ref{5.14}) yields
\begin{equation}\label{5.15}
a_n=\sum_{k=0}^n~\varrho^{2k}~ c_{k,n}(d)~\bar{a}_{n-k},~~~
b_n=\sum_{k=0}^n~\varrho^{2k}~ c_{k,n}(d-1)~\bar{b}_{n-k},
\end{equation}
\begin{equation}\label{5.16}
c_{k,n}(d)=\sum_{l=0}^k (-1)^l~{2^{2l} \over (2l)!}~
{\Gamma(l+n-d/2) \over \Gamma(n-d/2)}~ {1 \over (k-l)!}.
\end{equation}
The following property of $\Gamma$--functions \cite{BE}
\begin{equation}\label{5.17}
\sum_{n=-\infty}^{\infty}{\Gamma(a+n) \Gamma(b+n) \over
\Gamma(c+n) \Gamma (d+n)}=
{\pi^2 \over \sin(\pi a) \sin(\pi b)}~{\Gamma(c+d-a-b-1)
\over \Gamma(c-a)\Gamma(d-a)\Gamma(c-b)\Gamma(d-b)}
\end{equation}
($\mathrm{Re}~(a+b-c-d)<-1$ and $a,b$ are not integers)
can be used to show that $c_{k,n}(d)$ in (\ref{5.16}) coincide with
$c_{k,n-\alpha}(d)$ in (\ref{5.10}) at $\alpha=0$. This proves validity of
(\ref{3.17}), (\ref{3.18}).

\subsection{Spin 0 fields in rotating Einstein universe}

Consider a field
theory in the Einstein universe
$\mathrm{R}^1\times S^{3}$, where $S^3$ is a hyper-sphere of the radius $\rho$.
Because the Einstein universe is spatially compact
it allows globally defined frames of reference
which rigidly rotate with coordinate angular velocities $\Omega$, provided if
$\Omega<1/\rho$. In such frames, spectrum
of frequencies $\omega$ of single-particle field excitations
$e^{-i\omega x^0}\phi_\omega(x)$ is determined by a NLSP.
As a simplest example, consider conformally coupled scalar fields
\begin{equation}\label{5.18}
\left(-\nabla^2+\frac 16 R\right)\phi=0,
\end{equation}
where $R$ is the scalar
curvature of the universe, $R=6/\rho^2$.
In what follows we put $\rho=1$ and write
the metric in $\mathrm{R}^1\times S^{3}$ as
\begin{equation}\label{5.19}
ds^2=-(dx^0)^2+\sin^2\theta ~d\varphi^2+\cos^2\theta~ d\psi^2+ d\theta^2,
\end{equation}
where $0\leq \theta \leq \pi/2$, $0\leq \varphi,\psi \leq 2\pi$.
The metric in a frame which rotates with angular
velocity $\Omega$ can be obtained from (\ref{5.19}) by change $\varphi$ to
$\varphi+\Omega x^0$
\begin{equation}\label{5.20}
ds^2=-B(dx^0+a_\varphi d \varphi)^2+ {\sin^2\theta \over B}~
d\varphi^2+\cos^2\theta~ d\psi^2+d\theta^2,
\end{equation}
\begin{equation}\label{5.21}
B=1-\Omega^2\sin^2\theta,~a_\varphi=\Omega\sin^2\theta ~B^{-1}.
\end{equation}
To bring the corresponding NLSP to required form (\ref{1.3}),
(\ref{def}) we use conformal covariance and
first consider (\ref{5.18}) on a space with metric
related to (\ref{5.20}) by conformal transformation
\begin{equation}\label{5.22}
ds^2=-(dx^0+a_\varphi d \varphi)^2+
dl^2,
\end{equation}
\begin{equation}\label{5.23}
dl^2={1 \over B}\left[{\sin^2\theta \over B}~
d\varphi^2+\cos^2\theta~ d\psi^2+d\theta^2\right]=h_{jk}dx^j dx^k.
\end{equation}
Element $dl^2$ defines the metric on a compact three--dimensional
manifold ${\cal M}_3$ without boundaries.
The wave-equation on such a space-time
results to NLSP (\ref{1.3})
with operator\footnote{NLSP which appear
on stationary space-times are discussed in
\cite{Fursaev:2000},\cite{Fursaev:01} where further details
can be found.}
\begin{equation}\label{5.24}
L(\omega)=-(\nabla^k+i\omega a^k)(\nabla_k+i\omega a_k)+\frac 16 \bar{R}+
{1 \over 24} F^{jk}F_{jk}.
\end{equation}
Here $F_{jk}=a_{k,j}-a_{j,k}$ and $a_j dx^j=a_\varphi d\varphi$,
$\nabla_k$ are the covariant derivatives on ${\cal M}_3$,
$\bar{R}$ is the scalar curvature of ${\cal M}_3$.
Operator (\ref{5.24}) has form (\ref{def}).

The spectrum of the given NLSP can be easily found.
If $\omega_n$ are energies of quanta in the non-rotating frame (\ref{5.19})
the
energies in the rotating frame (\ref{5.20}) are $\omega_{nm}=\omega_n+m\Omega$
where $m$ is the projection of the angular momentum on the rotation axis.
For model (\ref{5.18}) the spectrum is $\omega_n=n+1$,
where $n=0,1,..$, which
follows from the spectrum of the Laplacian on $S^3$.
The number $m$ takes values $-n\leq m \leq n$ and $\omega_{nm}$
have degeneracy $d_{nm}=n-|m|+1$ for given $m$ and $n$, see, e.g. \cite{HR}.
The total spectrum of the NLSP also includes negative energies
$\omega_{nm}=-\omega_n+m\Omega$.
The positive (negative) energy states have
positive (negative) norm defined with respect to product (\ref{2.1}).
(The latter property is easy to understand if we note that signs of
$\omega_{nm}$ and $\omega_n$ are the same and (\ref{2.1}) coincides with
the Klein-Gordon product.)
Thus, requirement (ii) of section 2 is satisfied.

Because positive and negative parts
of the spectrum are symmetric the pseudo-trace, Eq. (\ref{1.4}),
is
\begin{equation}\label{5.25}
K(t)=\sum_{n=0}^\infty\sum_{m=-n}^{n}(n-|m|+1)~e^{-(n+1+m \Omega)^2t},
\end{equation}
Its short $t$ expansion should have form (\ref{3.16}) where
$d=3$ and $b_n=0$, because ${\cal M}_3$ has no boundaries.
The first coefficients $a_n$ for
(\ref{5.25}) can be found explicitly. For instance,
\begin{equation}\label{5.26}
a_0={2\pi^2 \over 1-\Omega^2}~,~~a_1=-{2\pi^2 \over 3}{\Omega^2 \over
1-\Omega^2}.
\end{equation}
The easiest way to get (\ref{5.26}) is to use the generalized
$\zeta$-function  $\zeta(\nu)=\sum_{nm}
d_{nm}\omega_{nm}^{-2\nu}$.
Then $a_0$ is the limit
$4\pi^2 (\nu-3/2)\zeta(\nu)$ at $\nu=3/2$ and
$a_1$ is $8\pi^2(\nu-1/2)\zeta(\nu)$ at $\nu=1/2$.

Expressions (\ref{5.26}) are in agreement with
formula (\ref{3.17}).
Indeed, one can check that $a_0$ is the volume of ${\cal M}_3$, see (\ref{5.23}).
For operator (\ref{5.24}) definitions (\ref{3.3b}) yield
\begin{equation}\label{5.27}
a_{0,1}=-{1 \over 24} \int_{{\cal M}_3}h^{1/2} d^3x~
F^{jk}F_{jk},~~ a_{2,2}=-{1 \over 12} \int_{{\cal M}_3}h^{1/2} d^3x~
F^{jk}F_{jk}.
\end{equation}
Coefficient $a_1$, as defined by (\ref{3.17}), is
\begin{equation}\label{5.28}
a_1=a_{0,1}+\frac 12 a_{2,2}=-{1 \over 12} \int_{{\cal M}_3}h^{1/2} d^3x~
F^{jk}F_{jk}
\end{equation}
and it coincides exactly with (\ref{5.26}).

\bigskip

\section{Spectral problems related to Dirac equation}
\setcounter{equation}0

In this section we consider spectral problems which follow
from the Dirac equation in a four-dimensional space-time
\begin{equation}\label{5.29}
\left[\gamma^\mu(\nabla_\mu-ie A_\mu)+M\right]\psi=0.
\end{equation}
Here $\nabla_\mu$ are spinor connections,
$A_\mu$ is a gauge potential and $M$ is a constant or a function, $M>0$.
Suppose that gravitational and
gauge background fields are stationary, i.e. there is a coordinate system
where $A_\mu$ and components of the metric do not depend on time $x^0$.
Without loss of generality we take the space-time metric in the form
\begin{equation}\label{s1}
ds^2=-(dx^0+a_j d x^j)^2+
dl^2,
\end{equation}
\begin{equation}\label{s2}
dl^2=h_{jk}dx^j dx^k,~~~j,k=1,2,3,
\end{equation}
where $a_j$ and $h_{jk}$  do not depend on $x^0$.
One can always make a conformal transformation in
(\ref{5.29}) to bring a stationary metric to this form.
We assume that (\ref{s1}) is the metric on a Riemannian manifold
${\cal M}_3$ and ${\cal M}_3$ is compact.
Let us use the basis of one-forms associated
with metric  (\ref{s1}), $U^0_\mu
dx^\mu=-dx^0-a_i dx^i$, $U^a_\mu U^a_\nu=h_{\mu\nu}$, $a=1,2,3$.
Then for single-particle
modes $\psi_\omega(x^0,x)=
e^{-i\omega x^0}\psi_\omega(x)$ the Dirac equation
is reduced to a Schr\"odinger equation
\begin{equation}\label{5.30}
(\omega-H(\omega))\psi_\omega=0,
\end{equation}
\begin{equation}\label{5.31}
H(\omega)=\breve{\gamma}_0(\gamma^k (D_k(\omega)-ie A_k)+M)+\frac i8 F
-\omega A_0~.
\end{equation}
Here $D_k(\omega)=\nabla_k+i\omega a_k$, $\nabla_k$ are spinor connections
on ${\cal M}_3$.
We define  $\gamma$--matrices
$\{\gamma_k,\gamma_j\}=2h_{kj}$ and the matrix  $\breve{\gamma}_0$
which anti-commutes with $\gamma_k$, $\breve{\gamma}_0^2=1$.
Also, $F=\gamma^j\gamma^k F_{jk}$ and $F_{jk}=a_{k,j}-a_{j,k}$.

Operator $H(\omega)$ is a Hermitian operator in a
Hilbert space $\mathrm{L}^2({\cal M}_3)$. If
$\lambda(\omega)$ are eigen-values of $H(\omega)$
the eigen-value problem (\ref{5.30}) for $\omega$
is reduced to equation
\begin{equation}\label{s3}
\omega-\lambda(\omega)=0.
\end{equation}
By using matrix $\gamma_5$, such that
$(\gamma^5)^2=1$ and it anti-commutes
with other $\gamma$'s, let us define a quadratic operator
polynomial with the same spectrum as the spectrum  of the Dirac problem
(\ref{5.30})
\begin{equation}\label{5.32}
-\left(\gamma_5 \breve{\gamma}_0~(\omega-H(\omega))\right)^2\phi_\omega=
(\omega^2-L(\omega))\phi_\omega=0,
\end{equation}
\begin{equation}\label{5.33}
L(\omega)=
-(D_k(\omega)-ieA_k-ib_k)(D^k(\omega)-ieA^k-ib^k)-\left(2e A_0+\frac i4 F\right)\omega +V,
\end{equation}
\begin{equation}\label{5.34}
b_k=\frac 14\breve{\gamma}_0~ F_{kj}~\gamma^j,
\end{equation}
\begin{equation}\label{5.35}
V=M^2-\gamma^k M_{,k}+\frac 14 \bar{R}
+{1 \over 16} F^{jk}F_{jk}
-\left(eA_0+\frac i8F\right)^2+
{e \over 2} ({\cal F}-2\breve{\gamma}^0\gamma^k\nabla_k A_0)~.
\end{equation}
Here $\bar{R}$ is the curvature of ${\cal M}_3$,
${\cal F}=2\nabla_jA_k\gamma^j\gamma^k$.
An alternative way to come to (\ref{5.32})
is to start with equation
$(\gamma^\mu (\nabla_\mu-ieA_\mu)-M)(\gamma^\nu(\nabla_\nu-ieA_\nu)+M)\phi=0$.

To define a quadratic operator polynomial for the Dirac problem (\ref{5.30})
one could also consider equation $(\omega-H(\omega))^2\phi_\omega=0$.
The difficulty is that it results in NLSP with the operator
$L(\omega)=-H^2(\omega)+2\omega H(\omega)$
which is not positive-definite. Another suggestion could be equation
$(\omega-H(\omega))(\omega+H(\omega))\phi_\omega=(\omega^2-H^2(\omega))\phi_\omega=0$.
However, it includes
eigen-values $\omega$ which are not eigen-values of the original problem (\ref{5.30}).

Operator $L(\omega)$ defined in (\ref{5.33}) has form (\ref{def})
and we assume that it is an elliptic operator with positive-definite
leading symbol (the case which is relevant for physical applications).
What differs $L(\omega)$ from the class of operators considered in section
2 is that it is not
Hermitian operator. The problem is that $b_k$ in (\ref{5.33})
and the last term in $V$ in  (\ref{5.35})
are anti-Hermitian matrices. Instead, $L(\omega)$
is Hermitian with respect to the following inner product:
\begin{equation}\label{5.36}
((\phi,\psi))=\int_{{\cal M}_3}h^{1/2} d^3x~\phi^{+}\breve{\gamma}_0\psi,
\end{equation}
which
is not positive-definite due to $\breve{\gamma}_0$.
Yet, because $\breve{\gamma}_0$ is Hermitian, it is possible to show by using
(\ref{5.36}) that the spectrum $\Lambda(\omega)$ of $L(\omega)$ is real
and by using positivity of the leading symbol of $L(\omega)$ it is possible
to show that the spectrum is bounded from below.

Let us define for spinors $\phi(x^0,x^i)$, $\psi(x^0,x^i)$ the following product
\begin{equation}\label{s4a}
\langle\langle \phi,\psi \rangle\rangle=i((\phi,\dot{\psi}))-
i((\dot{\phi},\psi))-((\phi,L_1\psi))
-i((\phi,L_0\dot{\psi}))+i((\dot{\phi},L_0\psi)),
\end{equation}
It has the same form as product
(\ref{2.1}). The difference is that $(~,~)$ is replaced to
$((~,~))$.
It is easy to show that $\langle\langle~,~\rangle\rangle$ is time-independent
on solutions to (\ref{1.1}).
Also one can see that for eigen-functions of (\ref{5.32})
\begin{equation}\label{s4}
\langle\langle \phi_\omega,\psi_\sigma \rangle\rangle=\delta_{\omega \sigma}~\chi'(\omega)
((\phi_\omega,\psi_\omega)),
\end{equation}
where $\chi'(\omega)$ is the derivative $\partial_\omega
\chi(\omega,\Lambda)$ at roots of (\ref{2.3}) and $\chi(\omega,\Lambda)$
is defined by (\ref{2.4}). By assuming that the norms do not vanish
condition (ii) of section 2 can be generalized in the following way: the
signs of the norms $\langle\langle \phi_\omega,\phi_\omega \rangle\rangle$
and $((\phi_\omega,\phi_\omega))$ coincide for $\omega>0$ and they are opposite for
$\omega<0$. According to (\ref{s4}) this guarantees that
$\chi'(\omega)=\epsilon(\omega)|\chi'(\omega)|$, the same property we had in
section 2.

If the gauge field is absent, $A_\mu=0$, the Dirac equation is
invariant under the charge conjugation $\psi\rightarrow \psi^c$, where
$\psi^c=C\bar{\psi}^{T}$, $\bar{\psi}=\psi^{+}\breve{\gamma}_0$,
and $C\gamma_\mu^{T} C^{-1}=-\gamma_\mu$.
It is easy to see that in this case the
spectrum of $\omega$ is symmetric, i.e., if $\psi_\omega$ is an eigen-function of
(\ref{5.32}), $\psi_\omega^{c}$ is an eigen function of the same problem
for the spectral parameter $-\omega$. In this case eigen-values of $L(\omega)$
are symmetric functions of $\omega$, which simplifies verification
of condition (ii).

Let us emphasize that $\langle\langle~,~\rangle\rangle$
does not coincide with the physical relativistic inner product $\langle~,~\rangle$
for Dirac equation (\ref{5.29})
\begin{equation}\label{s5}
\langle\phi,\psi\rangle=(\phi,\psi)-(\phi,H_0\psi),
\end{equation}
where $H_0$ is defined by the Hamiltonian $H(\omega)=H_1+\omega H_0$.
If $\phi$ and $\psi$ are solutions to (\ref{5.29}) then (\ref{s5}) does
not depend on $x^0$. If $\phi_\omega$ and $\psi_\sigma$ are solutions to
the Dirac problem (\ref{5.30})
\begin{equation}\label{s6}
\langle\phi_\omega,\psi_\sigma\rangle=\delta_{\omega\sigma}~\bar{\chi}'(\omega)
(\phi_\omega,\psi_\omega),
\end{equation}
where $\bar{\chi}'$ is the derivative
$\partial_\omega(\omega-\lambda(\omega))$ taken at the root of equation
(\ref{s3}). Because each solution $\psi_\omega$ to (\ref{5.30}) also is a solution to
(\ref{5.32}) there is a relation between two norms. One can show that
\begin{equation}\label{s7}
\langle\langle\psi_\omega,\psi_\omega\rangle\rangle
=\omega^{-1}\langle M \psi_\omega,\psi_\omega\rangle.
\end{equation}
Thus, if $M=0$ there are solutions of (\ref{5.32}) with zero norm determined with respect to
(\ref{s4a}). If $M$ is a positive constant and $\langle\psi_\omega,\psi_\omega\rangle$ is
positive it follows from (\ref{s7}) that
$\langle\langle\psi_\omega,\psi_\omega\rangle\rangle$ is positive for
$\omega>0$ and negative for $\omega<0$.

Consider now the pseudo-trace for the Dirac problem which we define
as earlier, as
\begin{equation}\label{s8}
K(t)=\frac 12 \sum_{\omega} e^{-t\omega^2},
\end{equation}
where $t>0$ and $\omega$ are eigen-values of quadratic polynomial
(\ref{5.32}). If operator $L(\omega)$, Eq. (\ref{5.33}) obeys the properties
formulated above in this section, $K(t)$ is related to the spectral density of
$L(\omega)$ by formulas (\ref{2.12})--(\ref{2.14}). These formulas can be
used to get for (\ref{s8}) asymptotic expansion (\ref{1.5}) where
coefficients are determined by coefficients of the asymptotic expansion of
the trace of the operator $e^{-tL(\omega)}$ by formulas (\ref{3.17}), (\ref{3.18}). In
other words, the results of this paper concerning asymptotic expansion
of $K(t)$ should be valid for spin 1/2 operators.
Note that operator $e^{-tL(\omega)}$ exists if $L(\omega)$ has a
positive-definite leading symbol regardless of the fact that $L(\omega)$ may
be a not self-adjoint operator \cite{Gilkey}, as it happens in the considered
case.

To illustrate these results let us return to the problem of
quantum fields in the rotating Einstein universe, see section 5.2.
The spectrum of $\omega$ in (\ref{5.30})
can be connected with the spectrum of the Dirac operator on
$S^{3}$.
For simplicity we consider massless neutral fields ($M=0$, $A_\mu=0$).
Then for Weyl spinors one gets (see \cite{HR})
$\omega_{nm}=\pm \omega_n+m\Omega $, where $\omega_n=3/2+n$, $n=0,1,2,...$,
and $-(n+1/2)\leq m\leq n+1/2$. The degeneracy of $\omega_{nm}$
is $d_{nm}=n-|m|+3/2$. The pseudo-trace for this spectrum is
\begin{equation}\label{5.37}
K(t)=
2\sum_{n=0}^\infty\sum_{m=-n-1/2}^{n+1/2}(n-|m|+3/2)~e^{-(n+3/2+m
\Omega)^2t}.
\end{equation}
The short $t$ expansion for (\ref{5.37}) is given by (\ref{3.16}) with
$d=3$, $b_n=0$. The first coefficients can be found explicitly
\begin{equation}\label{5.38}
a_0={4\pi^2 \over 1-\Omega^2}~,~~a_1={2\pi^2 \over 3}~{\Omega^2 -3 \over
1-\Omega^2}.
\end{equation}
To get the same result by our method one has to start with $M>0$ and then
go to the limit $M=0$.
This yields $a_0=2\mbox{vol}[{\cal M}_3]$ (factor 2 is related to
dimensionality of the spinor representation) and agrees with
$a_0$ in (\ref{5.38}).
Coefficient $a_1$ is determined by (\ref{3.17}),
(\ref{5.33}).  For Weyl spinors ($M=0$, $A_\mu=0$)
\begin{equation}\label{5.39}
a_{0,1}=\int_{{\cal M}_3}h^{1/2} d^3x~\mbox{Tr}\left[{1 \over
6}\bar{R}-V\right]=
-\int_{{\cal M}_3}h^{1/2} d^3x~\left[{1 \over
6}\bar{R}+{1 \over
16}F^{jk}F_{jk}\right],
\end{equation}
\begin{equation}\label{5.40}
a_{2,2}=\int_{{\cal M}_3}h^{1/2} d^3x~\mbox{Tr}\left[-{1 \over 32} F^2-{1
\over 12} F^{jk}F_{jk}\right]=-{1 \over 24}
\int_{{\cal M}_3}h^{1/2} d^3x~F^{jk}F_{jk}
\end{equation}
\begin{equation}\label{5.41}
a_1=a_{0,1}+\frac 12 a_{2,2}=-\int_{{\cal M}_3}h^{1/2} d^3x~
\left[\frac 16\bar{R}+{1 \over 12} F^{jk}F_{jk}\right],
\end{equation}
where $\bar{R}$ is the curvature of ${\cal M}_3$, see (\ref{5.23}).
By using definitions (\ref{5.21}), (\ref{5.23}) one can check that
(\ref{5.41}) coincides with $a_1$ in (\ref{5.38}).

\bigskip

Spectral asymptotics for spin 1/2 fields were studied in
\cite{Fursaev:2000}. In this work there is a mistake in definition of
$H(\omega)$. Equations (\ref{5.31}) and (\ref{5.41})
correct corresponding formulas of \cite{Fursaev:2000}.

\bigskip
\vspace{12pt} {\bf Acknowledgements}:\ \
I am grateful to V. Spiridonov for pointing out the book
\cite{Markus} and S. Zerbini for bringing my attention to the
problem discussed in section 5.1. I would also like to thank I. Avramidi,
G. Cognola, G. Esposito, S. Fulling, P. Gilkey, V. Moretti, and D. Vassilevich
for helpful discussions during preparation of this work.
This work is supported in part by the RFBR grant N 01-02-16791.

\newpage
\appendix
\section{Spectral asymptotics}
\setcounter{equation}0

The aim of this Appendix is to demonstrate that the
asymptotics (\ref{3.4}) hold when the corresponding operator has a number
of negative eigen-values. In this case we use definition (\ref{3.1})
\begin{equation}\label{A.1}
K_\omega(t)t^{-\alpha}=\int_{\mu}^{\infty} e^{-t\lambda} \varphi_\alpha (\lambda,\omega)
d\lambda,
\end{equation}
where parameter $\mu$ is smaller than the lowest eigen-value and, hence,
$\mu<0$. As earlier, we
suppose that $\alpha$ is a complex parameter, $\mathrm{Re}~\alpha >0$.
Equation (\ref{A.1}) is equivalent to the following equation
\begin{equation}\label{A.2}
\bar{K}_\omega(t)t^{-\alpha}=\int_{0}^{\infty} e^{-t\lambda}
\bar{\varphi}_\alpha (\lambda,\omega)
d\lambda,
\end{equation}
\begin{equation}\label{A.3}
\bar{K}_\omega(t)=e^{t\mu}K_\omega(t),
\end{equation}
\begin{equation}\label{A.4}
\bar{\varphi}_\alpha (\lambda,\omega)=\varphi_\alpha(\lambda+\mu,\omega)~.
\end{equation}
If $K_\omega(t)$ has asymptotics (\ref{3.3}) then
\begin{equation}\label{A.5}
\bar{K}_\omega(t) \sim {1 \over (4\pi t)^{d/2}}\sum_{n=0}^\infty
\left[\bar{a}_{n}(\omega)t^{n} +\bar{b}_{n}(\omega)t^{n+1/2}\right],
\end{equation}
and, according to (\ref{A.3}), the coefficients in the both
expansions are related as
\begin{equation}\label{A.6}
a_n(\omega)=\sum_{p=0}^n(-1)^p{\mu^p \over p!}\bar{a}_n,(\omega)~~~
b_n(\omega)=\sum_{p=0}^n(-1)^p{\mu^p \over p!}\bar{b}_n(\omega).
\end{equation}
(\ref{A.5}) defines the asymptotic expansion
of $\bar{\varphi}_\alpha(\lambda,\omega)$ at large $\lambda$
\begin{equation}\label{A.7}
\bar{\varphi}_\alpha(\lambda,\omega) \sim {1
\over (4\pi)^{d/2}}\sum_{n=0}^\infty\left[
\bar{a}_{n}(\omega)~ { \lambda^{d/2-n+\alpha-1}
\over \Gamma\left(\frac d2 -n +\alpha\right)}
+\bar{b}_{n}(\omega)~ { \lambda^{(d-1)/2-n+\alpha-1}
\over \Gamma\left({d-1 \over 2} -n +\alpha\right)}
\right].
\end{equation}
By using (\ref{A.4}) and (\ref{form}) we get from (\ref{A.7})
expansion which is valid for $\lambda>|\mu|$
\begin{equation}\label{A.8}
\varphi_\alpha(\lambda,\omega) \sim {1
\over (4\pi)^{d/2}}\sum_{n=0}^\infty\left[
a'_{n}(\omega)~ { \lambda^{d/2-n+\alpha-1}
\over \Gamma\left(\frac d2 -n +\alpha\right)}
+b'_{n}(\omega)~ { \lambda^{(d-1)/2-n+\alpha-1}
\over \Gamma\left({d-1 \over 2} -n +\alpha\right)}
\right],
\end{equation}
\begin{equation}\label{A.9}
a'_n(\omega)=\sum_{p=0}^n(-1)^p{\mu^p \over p!}\bar{a}_n(\omega),~~~
b'_n(\omega)=\sum_{p=0}^n(-1)^p{\mu^p \over p!}\bar{b}_n(\omega).
\end{equation}
After comparing  (\ref{A.9}) with (\ref{A.6}) one can replace
$a'_n(\omega)$, $b'_n(\omega)$
in (\ref{A.8}) to $a_n(\omega)$, $b_n(\omega)$, respectively.

\end{document}